\documentclass[preprint,prd,2showpacs,showkeys]{revtex4}
\usepackage{amsmath,graphicx}
       
\newcommand{\pni}{\par\noindent}

\begin{document}
\title{Wormhole solutions in the Randall-Sundrum 
scenario} 
\author{M. La Camera}
\email{lacamera@ge.infn.it} 
\affiliation{Department of Physics and INFN - University of Genoa\\  
Via Dodecaneso 33, 16146 Genova, Italy} 
\begin{abstract}
In the simplest form of the Randall-Sundrum model, we consider 
the metric generated by a static, spherically symmetric 
distribution of matter on the physical brane. The solution to the
five-dimensional Einstein equations, obtained numerically, 
describes a wormhole geometry. 
\end{abstract}
\maketitle
\newpage
\section{Introduction}
In the last few years large extra dimensions and brane worlds 
have been the subject of intensive investigations. The 
situation can be simplified to a five-dimensional problem where 
matter fields are confined to the four-dimensional spacetime 
while gravity acts in five dimensions. If the extra dimension has
a finite extension its compactification radius may be larger than
the Planck length without conflict with observations [1]. 
Alternatively, the extra dimension may be kept warped and
compactified or uncompactified [2,3] or even neither 
compact nor warped with a truly infinite size [4]. A considerable
effort has been devoted in exploring possible phenomenological 
and observable consequences of the brane world scenario. In 
particular Lorentzian wormholes [5,6],  smooth bridges 
connecting different universes or remote parts of the same 
universe, have gained much attention and  have  found a natural 
source in  brane world models  (see [7] and references quoted 
therein). In this letter we will show that a solution to Einstein
equations in the Randall-Sundrum scenario with a mass point on 
the physical brane can describe a wormhole geometry. The paper is
organized as follows. In Section II we start from the simplest  
form of the Randall-Sundrum model and modify the corresponding 
metric, which is asymptotically anti-de Sitter, to a metric 
generated by a static, spherically symmetric distribution of 
matter on the physical brane or, equivalently, axially symmetric 
in the bulk. By a suitable choice of coordinates we succeed in 
reducing the solution of the five-dimensional Einstein equations
to the solution of a single ordinary differential equation in one
unknown function. In Section III we solve numerically the highly 
nonlinear equation obtained and show that there are  wormhole 
solutions in the brane world. Some concluding remarks are given 
in Section IV. 
\section{The model}
There are in fact two models due to Randall and Sundrum with 
essentially the same framework. In the RS1 model [2] we have two 
3-branes with equal and opposite tensions which rigidly reside at
the boundaries ($y = 0$ and $y = r_c\pi$) of a slab of a 
five-dimensional anti-de Sitter bulk spacetime of radius $\ell$.
The five-dimensional Einstein equations in the vacuum are
\begin{equation}
G_{AB} = R_{AB} - \dfrac{1}{2} g_{AB} R = -\,\frac{\Lambda}{4 
M_5^3}g_{AB} 
\end{equation}
Here $M_5$ and $\Lambda$ denote 
respectively the five-dimensional Planck scale and the 
(negative) bulk cosmological constant. The Randall-Sundrum  
solution to the previous equations is
\begin{equation}
ds^2 = e^{-2 k\,|y|}\,\eta_{\mu\nu} dx^\mu dx^\nu + 
dy^2 
\end{equation}
where $\eta_{\mu \nu}$ is the flat Minkowskian metric, and $k = 
\frac{1}{\ell}$ is a scale factor for the extra dimension $y$. 
The above solution holds only if the tensions of the branes and 
the bulk cosmological term are fine-tuned  and in particular this
implies that $k$ must take the value $k = \sqrt{-\dfrac{\Lambda}{
24 M_5^3}}$. Moreover the following relation is derived
\begin{equation}
M_4^2 = \frac{M_5^3}{k}(1-e^{-2 k r_c \pi})
\end{equation}
which gives a well defined value for the 4D Planck scale $M_4$ 
even in the infinite radius limit. So if $M_5$ and $k$ are taken
both of order $M_4$ and $kr_c\pi \approx 35$ the model suggests a
solution to the hierarchy problem by showimg that TeV energy 
scales on the ``TeV brane'' at $y=r_c\pi$ correspond to 
$10^{16}$ TeV energy scales on the ``Planck brane'' at $y=0$. In 
an alternate scenario known as RS2 [3] the second brane, where 
now is supposed to be concentrated the graviton zero mode, is 
taken off at infinity letting $r_c \to \infty$, while the 
Standard Model particles are assumed to live on the brane at 
$y=0$. Such localization of gravity can obviate the need for the 
compactification of the fifth dimension. RS2 does not solve the 
hierarchy problem as RS1 does, but it is of interest from a 
purely gravitational point of view. While in the absence of 
matter the line interval can be exactly obtained, complications 
arise in the presence of matter fields, and even if an analytic 
expression could be found for the metric  produced by a  matter 
distribution on the physical brane it certainly would involve 
special functions, as one can see in linear and beyond linear 
order approximate solutions [8,9,10]. \pni Here we assume for 
simplicity that the matter fields reduce to a point mass on the 
physical brane. Having in mind the four-dimensional McVittie 
solution [11], which represents the gravitational field of a mass
particle in a Friedmann-Robertson-Walker universe and its 
generalization to higher dimensions [12], we employ  similar 
methods to obtain a  solution   although the metric we are 
searching  for is not spherically symmetric in five dimensions. 
\pni We start from the sourceless $AdS_5$ solution given by  
\begin{equation} 
ds^2 = e^{-2 k |y|}\,(dr^2 + r^2 d\Omega^2 -dt^2) + dy^2
\end{equation}
where $d\Omega^2$ is the line element on a two-dimensional sphere
of unit radius, and apply to it the following transformation of 
the Robertson type [13] 
\begin{equation}
\begin{cases}
r = & e^{k z}\,\dfrac{\rho}{\sqrt{1+k^2 \rho^2}} \\
{}& \\ 
y = & z-\dfrac{1}{2 k}\,\log{(1+k^2 
\rho^2)} 
\end{cases}
\end{equation}  
so obtaining
\begin{equation}
ds^2 = \dfrac{d\rho^2}{1+k^2 \rho^2} + 
\rho^2 d\Omega^2 - e^{-2k z}\,(1+k^2 \rho^2)\,dt^2 
+(1+k^2 \rho^2
)\,dz^2
\end{equation}
In the presence of a point mass $m$ bound at $r = 0$ to the 
physical brane in the old coordinates and consequently bound at 
$\rho = 0 $ in the new coordinates, we make the \emph{ansatz} 
that the line element (6) becomes
\begin{equation}
ds^2 = e^{\alpha(\rho)}\,d\rho^2 +  e^{\beta(\rho)}\,\rho^2 
d\Omega^2 -\, e^{-\, 2 k z + \gamma(\rho)}\,dt^2 +  
e^{\omega(\rho)}\,dz^2 
\end{equation}
In the above five-dimensional metric we used only four of the 
five available degrees of freedom, so (7) still has a  degree of 
generality which can applied in future contexts. From Einstein 
equations (1), written in the mixed form 
\begin{equation}
G_A^B = 6 k^2 \delta_A^B
\end{equation}
we obtain, omitting for simplicity the $\rho$-dependence 
\begin{equation}
G_t^z = \dfrac{k}{2} e^{\,\omega} (\gamma' - \omega') = 0 
\end{equation}
Therefore putting $\gamma = \omega$  one realizes by inspection 
that two of the components of the mixed Einstein tensor, namely 
$G_t^t$ and $G_z^z$, are equal.\pni Finally the relevant
equations are 
\begin{subequations}
\begin{eqnarray}
&&\dfrac{1}{4 \rho^2}\{e^{-\alpha}[4-4 e^{\alpha-\beta}+4 
e^{\alpha-\gamma} k^2 \rho^2 + \rho^2 {\beta'}^2 + 8 r \gamma' +
\rho^2 {\gamma'}^2  \nonumber\\
&&\hspace{0.3in}+ 4 \rho \beta'(1+\rho \gamma')]\} = 6 k^2 
\label{equationa}\\
&&\dfrac{1}{4 \rho}\{e^{-\alpha}[4 e^{\alpha - \gamma} k^2
\rho + \rho {\beta'}^2 + 4 \gamma'+3  \rho {\gamma'}^2 + 2 
\beta'(2 + \rho \gamma') \nonumber\\ 
&&\hspace{0.3in}- \alpha' (2 + \rho \beta' + 2 \rho \gamma') + 
2  \rho \beta'' + 4 \rho \gamma'']\} = 6 k^2
\label{equationb}\\
&&\dfrac{1}{4 \rho^2}\{e^{-\alpha}[-4 e^{\alpha - \beta} +
4  + 3  \rho^2 {\beta'}^2 + 4 \rho \gamma'+
\rho^2 {\gamma'}^2 + 2  \rho \beta' (6 +\rho \gamma') 
\nonumber\\ 
&&\hspace{0.3in}- \rho \alpha' (4 + 2 \rho 
\beta'+  \rho \gamma') + 4 \rho^2 \beta'' + 2
\rho^2 \gamma'']\} = 6 k^2  
\label{equationc} 
\end{eqnarray}\end{subequations}
Solving for $e^\alpha$ in equation (10a) and substituting this 
value into  equations (10b) and (10c) we obtain:
\begin{subequations}
\begin{eqnarray}
& e^{\alpha} = \dfrac{e^{\beta + \gamma}\,\left[\left( 2+\rho\, 
\beta'\right)\,\left(2+\rho\,\beta'+ 4 \rho\,\gamma'\right)+ 
\rho^2{\gamma'}^2\right]}{4\, \left[ e^\gamma + 
e^\beta\,\left(-\,1 + 6\,e^\gamma\right)\,k^2\rho^2\right]}
\label{equationd}\\
&\dfrac{3 \gamma'}{\left[\left( 2+\rho\, 
\beta'\right)\,\left(2+\rho\,\beta'+ 4 \rho\,\gamma'\right)+ 
\rho^2{\gamma'}^2\right]}\,\Phi(\rho) = 0
\label{equatione}\\
&\dfrac{3 (2+\rho \beta')}{\left[\left( 2+\rho\, 
\beta'\right)\,\left(2+\rho\,\beta'+ 4 \rho\,\gamma'\right)+ 
\rho^2{\gamma'}^2\right]}\,\Phi(\rho) = 0
\label{equationf} 
\end{eqnarray}\end{subequations} 
where
\begin{eqnarray}  
&&\Phi(\rho) = \,\, e^{-\beta}\rho\{\gamma'[\rho[\rho 
{\beta'}^2 + 4 (\rho \gamma' + 1)\beta'+ \gamma'(\rho
\gamma' + 8) - 2 \rho \beta'' ] + 8] \nonumber\\ 
&&\hspace{0.6in} + 2 \rho (\rho \beta' + 2) \gamma''\} - k^2 
\rho^2\{ 4 [\rho[\rho^2 {\beta'}^3 + 3 \rho (\rho 
\gamma' + 2){\beta'}^2\nonumber\\
&&\hspace{0.6in} - 3[\rho[\gamma'(\rho \gamma' - 4) + \rho 
\gamma'' ] -4]\beta' - 
\gamma'[\rho[\gamma'(\rho \gamma' + 6)  - 3 \rho \beta''] - 6] 
\nonumber\\
&&\hspace{0.6in} - 6 \rho \gamma''] + 8] 
- e^{- \gamma} [\rho[\rho^2 {\beta'}^3 + 2 \rho (2 
\rho \gamma' + 3) {\beta'}^2  + [\rho[\gamma'(\rho \gamma' + 
16)\nonumber\\
&&\hspace{0.6in} - 2\rho \gamma''] + 12]\beta' + 2   
\gamma'[\rho(\gamma' + \rho \beta'')+ 6] - 4 \rho \gamma''] + 8 
]\} 
\end{eqnarray}
Equation (11b) can be satisfied by $\gamma = 0$, but 
in this case (11c) and (11a) become respectively $k^2 = 0$ and 
$e^\alpha = e^\beta (2+\rho \beta')^2$ so it is immediate to 
verify, passing to standard coordinates, that we would 
have a flat five-dimensional space solution without cosmological 
constant. In the same way equation (11c) can be satisfied by 
$\beta=2 \ln(\rho_0/\rho)$ but, apart the unpleasant fact that 
the standard radial coordinate should have the constant value 
$\rho_0$, (11b) becomes $1+{\rho_0}^2k^2 = 0$  so cannot be 
satisfied. Finally the choice $\gamma = 0$ and $\beta=2 
\ln(\rho_0/\rho)$ is to be ruled out because it makes not soluble
already the system (10) from which we started. As a consequence 
(11b) and (11c) are both satisfied only by $\Phi(\rho)=0$. The 
fact that now we  are faced with the system (11a) and (12) of two
equations in the three unknown functions $\alpha, \beta, \gamma$ 
is not surprising because in writing the line element (7) we 
retained the possibility, due to the coordinate freedom of 
Einstein equations, of still applying one degree of freedom so 
we can arbitrarily reduce the unknown quantities from three to 
two. A possible choice can be made exploiting the solutions to 
the system in some limiting cases. If $m = 0$ and $k > 0$ the 
solution is given by the line element (6), while if $m > 0$ and 
$k = 0$ the solution has been investigated by various authors 
[14-17] and the corresponding line element, written in 
isotropic form, is given by 
\begin{eqnarray}
&&ds^{2} =\left( 1+\dfrac{\sqrt{3} m}{2\rho} \right)^{4} 
{\left| \dfrac{1-\dfrac{\sqrt{3} m}{2\rho}}
{1+ \dfrac{\sqrt{3} m}{2\rho}} \right|}
^{2\,\left( 1-2/\sqrt{3}\right)} \hspace{ 
-0.15in}(d\rho^{2}+ \rho^{2}d\Omega^{2})-\, 
{\left| \dfrac{1-\dfrac{\sqrt{3} m}{2\rho}}
{1+ \dfrac{\sqrt{3} m}{2\rho}} \right|}
^{2/\sqrt{3}}\hspace{ -0.1in}dt^{2}\nonumber\\
&&\hspace{0.3in}+{\left| \dfrac{ 1-\dfrac{\sqrt{3} m}{2\rho}}{1+ 
\dfrac{\sqrt{3} m}{2\rho}} \right|} ^{2/\sqrt{3}}dy^2 
\end{eqnarray}\raisetag{0.5in} 
The above line element was obtained in [17] as a solution to the 
five-dimensional Einstein equations outside a static and 
three-dimensional spherically symmetric distribution of mass $m$ 
and scalar charge $\sigma$ in equal amount and there it has 
been shown that it describes  a Lorentzian wormhole with throat
at $\rho = \sqrt{3}/2\,m$. Now comparing the $\rho$-dependence of
the coefficient multiplying $dt^2$ in the line elements (6) and 
(13), where respectively $m = 0$  and $k = 0$, a suitable but of 
course not the only one choice for the function 
$e^{\gamma(\rho)}$ when both $m$ and $k$ are different from zero 
may be the following: 
\begin{equation}
e^{\gamma(\rho)} =  {\left|  \dfrac{1-\dfrac{\sqrt{3} m} {2\rho}}
{1+ \dfrac{\sqrt{3} m}{2\rho}} \right|} ^{2/\sqrt{3}}(1 + k^2 
\rho^2) 
\end{equation}
So equation $\Phi(\rho)=0$ reduces to a second order differential
equation, which for simplicity is not written here, in the 
unknown function $\beta(\rho)$ and will be solved numerically. As
to the boundary conditions for $\beta$ and its first derivative, 
they will be fixed once we know the behavior of  $\beta(\rho)$ 
when $m/\rho$ is sufficiently smaller than unity and both $m$ and
$k$ are different from zero. This can be obtained from the 
equation $\Phi(\rho) = 0$ inserting in it the expansion of 
$\gamma(\rho)$ in those conditions and solving by series the 
resulting differential equation for $\beta(\rho)$. We find, with 
the requirement that $\beta$ vanishes at infinity 
\begin{equation} \beta = -\,\frac{2 
m}{\rho} -\, \frac{\left(\frac{m^3}{2} + \frac{2 
m}{k^2}\right)}{\rho^3} + \cdots 
\end{equation}
Opposite to the case $(m>0,k=0)$, where $\beta = 4 m/\rho + 
(3/2\, m^2)/\rho^2 + \cdots$, now $\beta$ reaches its asymptotic 
value $\beta = 0$ from below.
\section{Numerical calculations}
In view of searching for a numerical solution we have to 
fix numerical values for the two parameters $k$ and $m$. Now in 
both the models of Randall-Sundrum  one can derive from  equation
(3) the following relation between $k$ and the fundamental mass 
scales
\begin{equation}
k \sim \dfrac{M_5^3}{M_4^2}
\end{equation}
so with $M_5$ and $k$ taken, as in RS1, of order $M_4 \approx 
10^{19}\,GeV$, one obtains $k \approx 10^{32}\,mm^{-1}$ and 
since  $kr_c\pi \approx 35$ it follows that $r_c \approx 
10^{-31}\,mm$, so this is not a model with a large extra 
dimension. There are however various extensions of the 
generalized RS2 model (see e.g. [18] and references quoted 
therein) where the extra dimension is compact and much larger 
than Planck scale. In any case experiments impose the bound 
$r_c\alt 1\,mm$, which implies $k \agt 10 \, mm^{-1}$, so we 
shall let $k$ to vary by many orders of magnitude. In searching 
for the numerical solution to equation $\Phi(\rho) = 0$ we will 
pass to the adimensional variable $\rho/m$ and as a consequence 
in that equation $m$ and $k$ will appear only in the adimensional
quantity $mk$, but clearly not all the possible combinations of
$m$ and $k$ leading to the same value of $mk$ are physically 
acceptable. As a numerical example if $m$  refers to the
Sun mass then $m \approx 10^6 \,mm$ and therefore $10^7 \alt 
mk \alt 10^{38}$. The precision of calculations were 
tested verifying the reproduced value of the quantity $6k^2$ in 
the right-hand side members of Einstein equations (10b) and 
(10c) in a wide range of the variable $\rho$. The relative error 
proved to be a decreasing function of $\rho$, some of its 
approximate values being $10^{-2}$ at $\rho \alt m$, $10^{-4}$  
at $\rho = 2m$  and $10^{-6}$ at $\rho = 10^6\,m$. \pni To 
discuss the numerical solution of equation (12) let us consider 
the plots in figure 1. In the case $m > 0$ and $k = 0$ the 
function $e^{\beta}$, given now by equation (13), goes to 
infinity as $\rho$ goes to $\sqrt{3}/2\,m$ and decreases to one 
as $\rho \to \infty$. If $m = 0$ and $k > 0$ one has $e^{\beta} =
1$ from equation (6). Finally when both $m$ and $k$ are different
from zero $e^{\beta}$ becomes again infinite at $\rho = 
\sqrt{3}/2\,m$, reaches a minimum less than unity near $\rho = 
1.15\, m$  then increases towards the asymptotical value 
$e^{\beta} = 1$. It is worth noticing that the last behavior is 
numerically the same, in the limit of machine-precision, for all 
the values of $mk$ in the above considered range and this happens
because the values $mk \gg 1$ make equation $\Phi(\rho) = 0$ very
weakly dependent on $mk$. As to the function $e^\alpha$ given by 
equation (11a) we found, using the calculated values of $\beta$ 
and for all values of $mk$ as before, that it is positive 
when $\rho$ is greater than $\rho_{min} \approx 
(\sqrt{3}/2 + \varepsilon )\,m$, where $\varepsilon \approx 
2.5\,10^{-4}$. Our solution is therefore defined for $\rho \geq 
\rho_{min}$. At $\rho_{min}$ the functions $e^\beta$ and 
$e^\gamma$ are finite, different from zero and both much greater 
than unity. Finally, for $ \rho_{min} \leq \rho < \infty $ the 
function $e^{\alpha}$ results decreasing and of order $1/(k^2 
\rho^2)$, except in the interval $(0.95\,m,1.12\,m)$ where it is 
slightly increasing. 
\begin{figure} 
\begin{minipage}[t]{0.40\linewidth}      \centering
\includegraphics[width=1.0\linewidth]{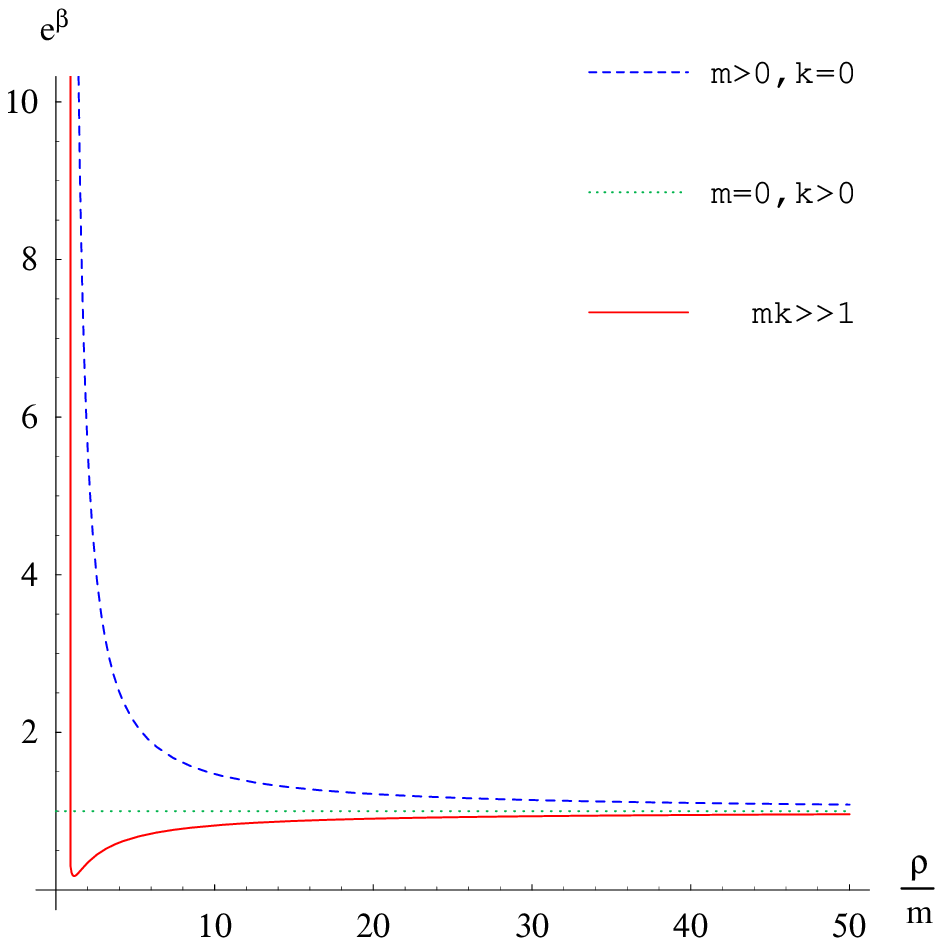}     
     \caption{The  function $e^{\beta}$ for different values of 
the pair (m,k).} 
\end{minipage}
\hspace{0.2in}
\begin{minipage}[t]{0.40\linewidth} 
    \centering
    \includegraphics[width=1.0\linewidth]{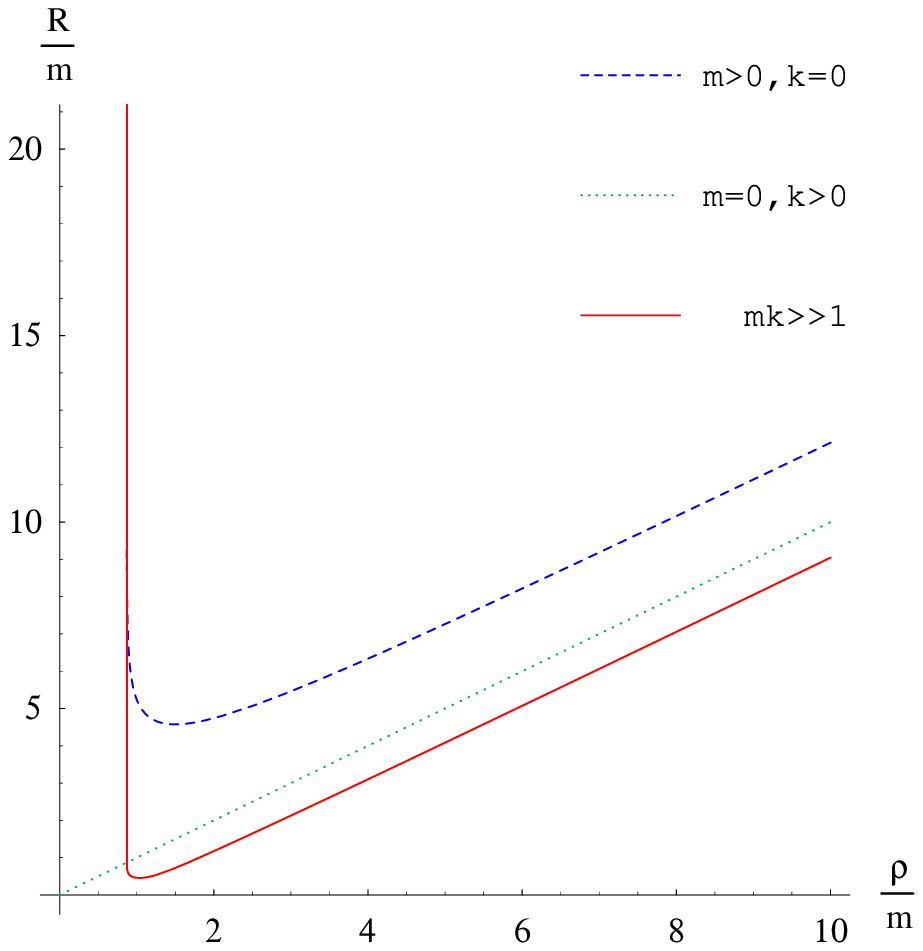}
    \caption{The function $\frac{\mathrm{R}}{\mathrm{m}}$ for 
different values of the pair (m,k).} 
\end{minipage}
\end{figure}
To see what is going on, look at the standard radial coordinate
$R(\rho) = \rho e^{\beta/2}$  depicted in figure 2 for  different
values of the pair $(m,k)$. The case $mk \gg 1$ we are 
treating here is similar to the case $(m > 0,k = 0)$ of  equation
(13) already discussed in [17] and will bring to analogous 
conclusions. More in detail, consider the proper circumference of
radius $\rho$ given by $C(\rho) = 2 \pi R(\rho)$. We have that 
$C(\rho)$ becomes infinitely large as $\rho \to \infty$ and 
asymptotically very large,  mantaining however finite, as $\rho 
\to \rho_{min}$ and  has a minimum at $\rho_0 \approx
1.04\,m$. This numerical value was estimated plotting the 
derivative $dR/d\rho$ versus $\rho$ and looking at the value of 
$\rho$ where it vanishes. The regions $\rho \in 
(\rho_{min},\rho_0)$ and $\rho \in (\rho_0,\infty)$ are two 
spacetimes associated with a  wormhole whose throat occurs at 
$\rho_0$ where $C(\rho)$ is minimum. The wormhole is  in this 
case aymmetric under the interchange of the asymptotic  surfaces 
($\rho = \rho_{min}$ and $\rho = \infty$) and traversable in 
principle due to the absence of event horizons. We verified 
numerically that the Kretchmann invariant becomes infinity only 
at $\rho = \sqrt{3}/2\, m$ which is not part of the manifold. 
\section{Conclusion} Some remarks seem here to be appropriate. 
First, it is known that a wormhole geometry can appear only if 
some of the energy conditions are violated and this requires the 
presence of some amount of ``exotic'' matter. The violation could
be justified incorporating the cosmological constant into an 
energy-momentum tensor describing the vacuum contribution to the 
density and pressure. However wormhole solutions can be obtained 
also with vanishing cosmological constant in a Ricci flat 
spacetime, where energy conditions are trivially satisfied,  of 
dimension $N \geq 5$ as in [17] where it was shown that every 
time an extra dimension is reduced a massless scalar field 
appears and plays the role of ``exotic'' matter. Also we would 
like to stress that in the framework of induced matter theory 
[19] the four-dimensional Einstein equations with sources can be 
locally embedded in five-dimensional Einstein equations without 
sources so although the matter seems exotic in four dimensions 
the five-dimensional spacetime  on both sides of the throat is 
that of a vacuum and energy conditions are again trivially 
satisfied. Second, according to the common assumption made both 
in brane world and in induced matter theories, our 
five-dimensional metric tensor depends explicitly on the fifth 
coordinate, as can best seen inverting transformation (5) and 
applying it to the line element (7). One obtains
\begin{eqnarray} 
&&\hspace{-0.5in}ds^2 =\, 
e^{-2k|y|}\,\Big\{\Big[e^{\overline{\alpha}} + 
\dfrac{e^{\overline{\gamma}}k^2r^2e^{-2k|y|}} {(1+ 
k^2r^2e^{-2k|y|})^2}\Big]dr^2+ e^{\overline{\beta}}r^2 d\Omega^2-
\dfrac{e^{\overline{\gamma}}}{(1+ 
k^2r^2e^{-2k|y|})}dt^2\Big\}\nonumber\\ 
&&\hspace{-0.5in}+\Big[e^{\overline{\alpha}} k^2r^2e^{-2k|y|} + 
\dfrac{e^{\overline{\gamma}}}{(1+ k^2r^2e^{-2k|y|})^2}\Big] dy^2+
2kr e^{-2k|y|}\Big[-e^{\overline{\alpha}} + 
\dfrac{e^{\overline{\gamma}}}{(1+ k^2r^2e^{-2k|y|})^2}\Big] drdy 
\end{eqnarray}
where $\overline{\alpha}$, $\overline{\beta}$ and  
$\overline{\gamma}$ denote the functions $\alpha$, $\beta$ and 
$\gamma$ written in the coordinates $r$ and $y$. In this short 
letter we have not considered more general gauge   
transformations. Further investigations should have to take into
account, as pointed out in [20], the influence that coordinate 
transformations may have on the interpretation of physical 
quantities in one lower dimension when in the reduced dimension 
is still present the extra coordinate.


\newpage 
\begin{thebibliography}{00} 
\bibitem{01} N. Arkani-Hamed, S. Dimopoulos, G. Dvali, Phys. 
Lett. B 429 (1998) 263 hep-ph/9803315; Phys. Rev. D 59 (1999) 
086004 hep-ph/9807344; I. Antoniadis,  N. Arkani-Hamed, S. 
Dimopoulos, G. Dvali, Phys. Lett. B 436 (1998) 257 
hep-ph/9804398. 
\bibitem{02} L. Randall, R. Sundrum, Phys. Rev. 
Lett. 83 (1999) 3370 hep-ph/9905221.  
\bibitem{03} L. Randall, R. Sundrum, Phys. Rev. Lett. 83 (1999) 
4690 hep-th/9906064. 
\bibitem{04} G. Dvali, G. Gabadadze, M. Porrati, Phys. 
Lett. B 485 (2000) 208 hep-th/0005016.
\bibitem{05} M. S. Morris, K. S. Thorne, Am. J. Phys. 56 (1988) 
395; M. S. Morris, K. S. Thorne, U. Yurtsever, Phys. Rev. Lett. 
61 (1988) 1449.
\bibitem{06} M. Visser, ``Lorentzian Wormholes: from Einstein to 
Hawking''(AIP, Woodbury, 1995).
\bibitem{07} K. A. Bronnikov, Sung-Won Kim, Phys. Rev. D67 (2003)
064027, gr-qc/0212112.
\bibitem{08} S. Giddings, E. Katz, L. Randall, JHEP 0003 (2000) 
023 hep-th/0002091.
\bibitem{09} I. Giannakis, H. C. Ren, Phys. Rev. D63 (2001) 
024001 hep-th/0007053. 
\bibitem{10} H. Kudoh, T. Tanaka, Phys. Rev. D64 (2001) 084022 
hep-th/0104049. 
\bibitem{11} G. C. McVittie, Mon. Not. R. Astron.
Soc. 933 (1933) 325.
\bibitem{12} L. K. Patel, R. Tikekar, N. Dadhich, Grav. Cosmol.  
6 (2000) 335 gr-qc/9909069.
\bibitem{13} H. P. Robertson, Phil. Mag. 5 (1928) 835.
\bibitem{14} A. Davidson, D. A. Owen, Phys. Lett. B 155 (1985) 
247. 
\bibitem{15} P. S. Wesson, Phys. Lett. B 276 (1992) 299. 
\bibitem{16} D. J. Gross, M. J. Perry, Nucl. Phys. B 226 (1993) 
29. 
\bibitem{17} A. G. Agnese, M. La Camera, Phys. Rev. D 58 (1998) 
087504 gr-qc/9806074.
\bibitem{18} R. Maartens, Prog. Theor. Phys. Suppl. 148 (2002) 
213, gr-qc/0304089. 
\bibitem{19} P. A. Wesson, ``Space-Time-Matter'' (World 
Scientific, Singapore, 1999).
\bibitem{20} J. Ponce de Leon, Int J. Mod. Phys. D 11 (2002) 1355
gr-qc/0105120.
\end{thebibliography}
\end{document}